# Homolytic Cleavage of Water on Magnesia Film Promoted by Interfacial Oxide-Metal Nanocomposite

Zhenjun Song[†] and Qiang Wang[‡]

[†] *Department of Chemistry, Key Laboratory of Advanced Energy Material Chemistry (MOE), and Collaborative Innovation Center of Chemical Science and Engineering (Tianjin), Nankai University, Tianjin 300071, People's Republic of China.*

[‡] *State Key Laboratory of Coal Conversion, Institute of Coal Chemistry, Chinese Academy of Sciences, Taiyuan 030001, Shanxi, People's Republic of China.*

**ABSTRACT:** The atomic-scale insights into the interaction of water with oxide surface are essential for elucidating the mechanism of physiochemical processes in various scientific and practical fields, since the water is ubiquitous and coats multifarious material surface under ambient conditions. By utilizing periodic density functional theory calculations with van der Waals corrections, herein we report for the first time the energetically and thermodynamically favorable homolytic dissociative adsorption behavior of water over magnesia (001) films deposited on metal substrate. The water adsorption on pristine magnesia (001) is quite weak and the heterolytic dissociation is the only fragmentation pathway, which is highly endothermic with large activation barrier of 1.167 eV. The binding strength for the molecular and dissociative adsorption configurations of water on MgO(001)/Mo(001) are significantly larger than corresponding configurations on bare magnesia (001). The homolytic dissociative adsorption energy of water on monolayer oxide is calculated to be -1.192 eV, which is even larger than all the heterolytic dissociative adsorption states. With the increase of the oxide thickness, the homolytic dissociative adsorption energy decreases promptly, indicating the nanoscale of oxide film play a crucial role in homolytic splitting of water. The homolytic dissociative adsorption structure could be obtained by transformation reaction from heterolytic dissociative adsorption structure, presenting activation barriers 0.733, 1.103, 1.306, 1.571, and 1.849 eV for reactions on 1 ML – 5 ML films. Bader charge population and differential charge density contours are analyzed to interpret charge transfer mechanism in the composite structure and between surface and adsorbates. The electronic states of dissociative adsorbates are located at lower relative energy ranges, with decreasing film thickness, which further implies the ultrathin oxide films are more favorable for stabilizing the produced hydride and hydroxyl species. The bonding characteristics between hydride/hydroxyl of homolytically dissociated water with ultrathin magnesia (001) are uncovered by characterizing the electron localization function and particular occupied orbitals. It is anticipated that the results here could provide inspiring clue and versatile strategy for enhancing chemical reactivity and properties of insulating oxide toward homolytic water-splitting processes involved in novel and unprecedented applications.

## 1. Introduction

Physiochemical process occurring on metal oxide plays an important role in surface science.[1] Surface defects change the local surface stoichiometry and influence the surface properties greatly. Functionalization of metal oxides with defects are extensively adopted for realizing high reactivity of oxide surfaces, and the color centers and vacancies are frequently involved in chemical reactions due to more attractive defect-adsorbate interaction compared to the perfect surface areas.[2] In addition to surface defects, the oxide surface with other unsaturated surface sites such as steps, kinks and corners can also enhance the catalytic reactivity and properties of metal oxides.[3] The embedded surface defects in well-defined oxide can donate or abstract electrons to adsorbed species, which promotes the activity of unreactive insulating magnesia considerably.[4] Metal particles dispersed over oxide substrates are extensively employed as a conventional way in industrial catalyst, and the supported metal atoms are directly involved in the catalytic reactions.[5] Compared with the bulk oxides, scientists know much less about the special structures and catalytic performance of oxide films at the nanoscale. Rodriguez et al investigated the unique properties of inverse deposition of nanometer ceria on metallic crystalline surface, and prospected the inspiring advances and opportunities of inverse oxide/metal catalyst.[6] The strong oxide-metal interaction (SOMI) present in inverse oxide/metal catalyst activates the supported oxide greatly through charge transfer, and the inverse catalysts show remarkable reactivity for a number of very important reactions such as CO oxidation,[6] water gas shift reaction,[7] $CO_2$ conversion,[8] and steam reforming of methane/alcohol.[9]

Magnesia, a representative ionic crystal, has become one of the crucial model systems for understanding the very sophisticated surface structures and phenomena. The adhesion behavior of metallic adsorbates (such as copper,[10] nickel,[11] palladium,[12] silver,[13] rhodium,[14] cobalt,[15] gold,[16] lithium,[17] potassium[18]) on ideal MgO (001) surface are extensively studied, which provide meritorious implications for understanding the growth patterns of metal clusters on inert magnesia and the special properties of interfacial metal-oxide composite structure. Apart from the metal/magnesia interfaces, the adsorption of metal nanoclusters and important molecules on metal-supported magnesia films becomes a topic of interest experimentally and theoretically to reveal unusual catalytic behavior of inert insulating oxide at nanoscale. The $Pd_n$ and $Rh_n$ clusters deposited on magnesia thin films epitaxially grown on metal substrates exhibit peculiar activity and selectivity for acetylene polymerization,[12b, 19] and show considerably strong binding energies toward CO.[20] On supported thin magnesia film, the equilibrium structure of small gold clusters with three to seven atoms prefer linear configuration, and the 2-dimensional gold plane prevails only for the even larger atom assemblies.[21] In addition, Pacchoini et al. confirmed unambiguously the formation of gold anions on ultrathin magnesia films by accurate analyses of the observable consequences obtained theoretically.[22] The electrostatic interaction between the metal substrate and the metal-induced excess charge accumulation at the adsorbate-oxide interface should be responsible for the enhanced bonding interaction between gold nanoclusters and magnesia.[23] Risse et al.[24] elucidated the adsorption behavior of gold dimers on thin magnesia film using low temperature scanning tunneling microscopy and density functional calculations, and found that the dimers can exist in both neutral upright structure and charged flat lying isomer, which implies microscopically that the electronic properties of adsorbates can vary considerably with adsorption geometries and sites. The adsorption site (above the oxide film or at the interface) and the coverage of alkali metal were found to be important factors which substantially lower the work function of oxide-metal hybrid systems.[25] For alkali atoms adsorbed on thin magnesia films, the preferred binding sites are surface oxide anions, with different anion locations as a function of the deposition temperature and metal categories, and the hyperfine coupling constants are reduced due to polarization effects.[26] Savio et al.[27] revealed by STM analysis and DFT calculations the formation of flat $Ni_xO_y$ aggregates facilitated by segregation of interfacial oxygen atoms, which give a further proof of the peculiar behavior of ultrathin oxide films. Tosoni et al.[28] investigated the bonding modes of $CO_2$ with bulk magnesia, Al-doped magnesia, and metal-supported magnesia films, and found that the activation effects and adsorption properties of ultrathin magnesia films strongly depend on the metal supports. Chen et al.[29] reported the heterolytic and homolytic dissociative splitting of dihydrogen on thin magnesia films, and deduced that the choice of the support and surface morphology are crucial for particular favored dissociation mode. Various adsorption modes originating from the interaction of methanol with the regular and defected thin magnesia films were studied by employing Fourier transform infrared and thermal desorption spectroscopies, and ab initio cluster/periodic model calculations,[30] and the results indicate the defect rich films should be highly reactive and responsible for the hydrogen release. $N_2O_2$ dimer forms preferentially on neutral F centers on ultrathin magnesia (001), and subsequently the reduced product $N_2O$ can be obtained with small activation barrier of about 0.1 eV.[31] The interaction of NO monomer with low-coordinated cations of thin magnesia is strong and prevents the formation of diamagnetic species, which is responsible for small fraction of paramagnetic chemisorbed species in EPR experiments.[32] Recently, Song et al.[30b, 33] uncovered the remarkably strong chemisorption of nitric oxide and the splitting behavior of molecular methanol on metal-supported ultrathin magnesia (001).

The adsorption and splitting behavior of water over metal oxides has attracted a great deal of attention because of its fundamental importance in electrochemistry, surface science, solar energy conversion, geochemistry and heterogeneous catalysis.[1a, 34] The interaction of water with magnesia has been extensively studied previously. Kim and Kawai's group reported two dissociation pathway (vibrational excitation and electronic excitation) of water on ultrathin magnesia film using low-temperature scanning tunneling microscopy.[35] The enhanced chemical activity for water dissociation was found on silver supported magnesia film, which possibly originates from the high stability of dissociative adsorbates and the strong hybridization of electronic states at the interface structure.[36] The strain-induced enhanced reactivity of ultrathin magnesia deposited on Mo(001) that causes water dissociation has been identified at the single-molecule level.[37] By coadsorption of water and oxygen, a series of highly reactive oxygen species including superoxide, hydroperoxide, hydroxyl and single oxygen adatoms are formed on metal-supported magnesia films.[38] The X-ray photoelectron and Auger spectroscopy also verified the greatly enhanced dissociation behavior of water on ultrathin magnesia film.[39] Rational manipulation of interfacial local structure by introducing vacancies and ligand field effect, which changes the charge distribution at the oxide surface, play important role in controlling the surface reactions.[40] Goodman et al. predicted that dissociative adsorption of $H_2O$ at the magnesia-water interface is energetically more favorable than molecular adsorption, due to polarization of the surrounding solvent.[41] The metastable impact electron spectroscopy and thermal programmed desorption reveal the prominent multilayer water adsorption before the entire coverage of water.[42] Compared with the terrace sites, the low-coordinated anions of MgO at step, edge and corner sites show much higher basicity and reactivity. Pacchioni and Freund discussed in detail the important influence of nanostructuring and nanodimensionality on the occurrence of electron transfer in the combined oxide films and metal substrate.[43] The deep understanding of electron transfer effect is essential to get atomic-scale information and reactive activity about magnesia, which is a wide gap insulator. The particular charge state and high thermal stability of adsorbates can be induced by adsorbate-surface interaction on metal-supported ultrathin oxide films.[5c, 16, 44] In this study, the homolytic fragmentation states of water on ultrathin magnesia films are investigated in detail. As far as we know, the energetically favorable homolytic fragmentation channel of water on perfect insulating oxide such as magnesia (001) has never been reported before without participation of surface defects or morphological irregularities.

## 2. Methodologies and Models

We perform periodic density functional theory calculations using gradient−corrected Perdew-Burke−Ernzerhof (PBE) exchange-correlation functional.[45] As the polarization and dispersion forces play an important role for the adsorption of water on the insulating magnesia and the adhesion of magnesia film to the metal support, the DFT−D3 (Version 3.0, Rev1) approach proposed by Grimme et al.[46], which for the first time introduces geometry dependent information in dispersion correction by employing the new concept of fractional coordination numbers to interpolate between dispersion coefficients of atoms in different chemical environments, is added on to the standard Kohn-Sham density functional theory to accurately consider the van der Waals (vdW) interactions. The projector augmented wave (PAW) technique[47], which further develops the US−PP concept by combining ideas from the pseudopotential and linear augmented−plane wave methods, is adopted to describe the interaction between the core and valence electrons. The plane wave basis sets are used to represent the valence electrons of the systems, and a large kinetic energy cutoff of 500 eV is used to expand the Kohn-Sham orbitals. The cutoff radii for calculating pair interactions and coordination number in vdW−energy expression are 50 Å and 20 Å, respectively. The damping function parameters for sixth and eighth order dispersion terms are 1 and 0.7875 respectively.

The lattice constants of bulk magnesia and molybdenum are optimized to be 3.121 Å and 4.201 Å respectively, which agree well with their experimental values 3.147 Å and 4.212 Å.[48] The metal substrate has been modeled by four molybdenum layers. During the calculations, we use the supercell consisting of 4×4 unit cells of molybdenum (001). The 1 layer – 5 layer ultrathin magnesia films has been deposited on the metal slab to construct the MgO(001)/Mo(001) hybrid structures. The supercells contain 16 Mg and 16 O, and 16 Mo atoms per layer. Due to the mismatch between the bulk magnesia and molybdenum, the magnesia surface are extended by 4.8% after deposition on molybdenum substrate. The single crystal surface of bulk magnesia (001) is represented by six layers magnesia slab, with the bottom two layers magnesia fixed to their bulk position. The bottom two layers of molybdenum slab are fixed to their bulk position to mimic structure and properties of inner bulk molybdenum, while top two layers molybdenum and the magnesia film of MgO(001)/Mo(001) composite surface are fully relaxed until all atomic Hellmann−Feynman forces are lower than 0.02 eV Å$^{-1}$. The (2×2×1) and (4×4×1) Gamma-Centered k-points methes are used to sample the first Brillouin zone, for the energy minimization during structural relaxation and the description of the electronic properties, respectively. The periodic boundary condition in $z$ direction with a large vacuum gap with the distance of 17 Å is applied to avoid image-image interaction and separate successive slabs. The fragmentation reaction profiles and activation barriers are obtained using the climbing image nudged elastic band (CI-NEB) method implemented in VTST code,[49] which finds the highest saddle point more efficiently and accurately than original method thanks to the better tangent definition and improved estimate of reaction coordinate. The aforementioned electronic structure calculations on the slab models are based on density functional theory at the level of generalized gradient approximation (GGA) and performed with Vienna *Ab Initio* Simulation Package (VASP).[50] The adsorption energies of molecular or dissociative water on bare magnesia (001) are obtained by the formula

$$E_{ad} = E(\text{water/MgO(001)}) - E(\text{water}) - E(\text{MgO(001)}) \quad (1)$$

The adsorption energies of molecular or dissociative water on metal-supported ultrathin magnesia (001) films are calculated as

$$E_{ad} = E(\text{water/MgO(001)/Mo(001)}) - E(\text{water}) - E(\text{MgO(001)/Mo(001)}) \quad (2)$$

The negative sign of adsorption energies ($E_{ad}$) corresponds to an exothermic adsorption or dissociation. Utilizing the charge density decomposition program developed by Henkelman et al.,[51] the Bader charge population is analyzed on the obtained total charge density, at very fine fast Fourier transform grids that accurately reproduce the correct total core charges. Electronic and geometric structures for adsorption and fragmentation of water on bare and metal-supported magnesia (001) are analyzed and visualized using the Visual Molecular Dynamics (VMD) program,[52] VASPMO program,[53] together with the VESTA program.[54]

## 3. Results and discussion

### 3.1 Water adsorption on pristine magnesia (001)

We first performed DFT-D3 calculation to illustrate the formation of molecular adsorption state (A) on bare magnesia (001) surface. The water molecules and the bare magnesia (001) slabs are preoptimized, and the favorable adsorption structure is obtained with one surface magnesium occupied by water molecule (Figure 1). The hydrogen of water interact with surface oxygen through hydrogen bonds. The binding energy of water on the bare magnesia (001) is calculated to be -0.653 eV at DFT-D3 level, and the dissociative adsorption energy is 0.504 eV. The dissociation of water is highly endothermic by 1.157 eV with large activation energy barrier 1.167 eV, leading to the formation of surface hydroxyl groups.

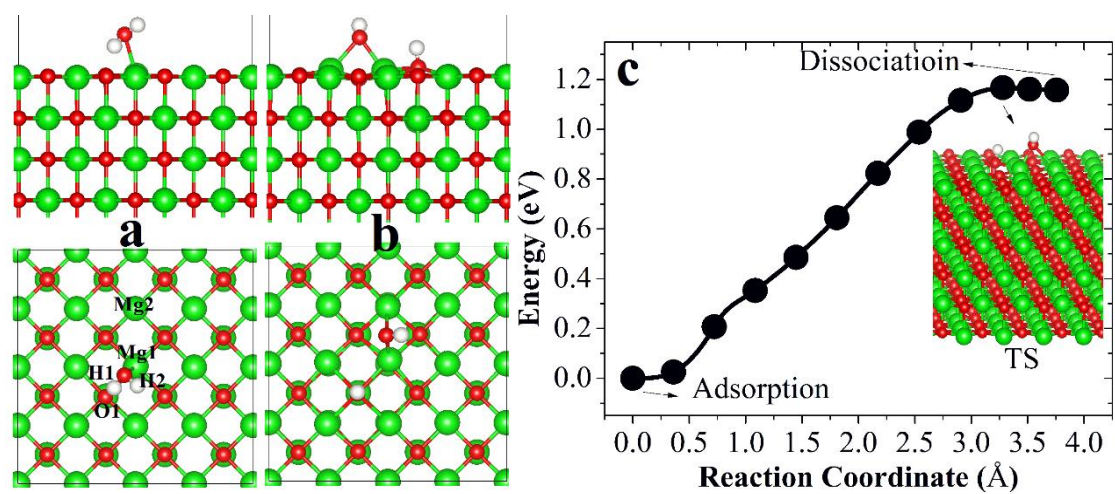

**Figure 1.** Geometric structures for water adsorption (a) and dissociation (b) on bare magnesia (001); (c) the reaction pathway for water dissociation on bare magnesia (001) at DFT-D3 theoretical level.

### 3.2 Water dissociation on metal-supported magnesia (001)

The binding strength for molecular adsorption of water on molybdenum supported magnesia (001) films are substantially larger than that on bare magnesia slab. Quantitatively, the adsorption energies for molecular water are calculated to be -0.805 eV, -0.853 eV, -0.826 eV, -0.854 eV and -0.832 eV on 1 ML ~ 5 ML ultrathin oxide films deposited on molybdenum, respectively. As listed in Table 1, the bond distances of $O_w$-H1 are in the range of 1.043 Å ~ 1.063 Å, which is slightly lengthened than that of isolated water molecule (0.972 Å). The O1-H1 distances are in the range of 1.502 Å ~ 1.536 Å, which can be typically classified as hydrogen bond. The molecular adsorption of water leads to surface rumpling of top layer. The monolayer oxide film deposited on molybdenum exhibits largest degree of surface rumpling. The adsorbed water molecules are negatively charged with the amount of charge between -0.076 e and -0.288 e. The charging amount of water on even layer oxide films are larger than that on odd layer oxide, essentially in agreement with the sequence of adsorption energies, as shown in Figure S1. The computed net charges on H1 are smaller when the hydrogen bonding interaction between H1 and surface oxygen are stronger (Figure S1). We have also derived the surface rumpling versus film thickness curves, which indicate that the rumpling values are more severe for the structures with an odd layer oxide films. The smaller surface rumpling facilitate the larger adsorption energy.

**Table 1. The geometric parameters and surface rumpling (in angstrom) for molecular water adsorption on 1 ML ~ 5 ML ultrathin oxide films deposited on molybdenum, at DFT-D3 theoretical level.**

| Film thickness | $O_w$-Mg | $O_w$-H1 | O1-H1 | $O_w$-H2 | $\Delta z^a$ |
|---|---|---|---|---|---|
| 1 ML | 2.098 | 1.043 | 1.536 | 0.972 | 0.309 |
| 2 ML | 2.109 | 1.063 | 1.502 | 0.972 | 0.117 |
| 3 ML | 2.116 | 1.053 | 1.525 | 0.971 | 0.136 |
| 4 ML | 2.112 | 1.052 | 1.530 | 0.972 | 0.123 |
| 5 ML | 2.117 | 1.051 | 1.533 | 0.972 | 0.125 |

[a] The surface rumpling are defined as the projected distance of the top atom and the bottom atom of the first layer, in $z$ direction.

Bader charge populations (in electron) of Ow, H1, H2, O1, Mg1 atoms, the magnesia films and molybdenum substrates for molecular water adsorption on 1 ML ~ 5 ML ultrathin oxide films deposited on molybdenum are calculated and listed in Table 2. The charge amounts of H2 are smaller than that of H1, due to the multiple ionic interaction between H1 and surface magnesium. The net charges of Ow exhibit mild even-odd alteration effect, which is similar to varying pattern of charges of water. As the O1 of 1 ML oxide film form ionic bonds with surrounding magnesium, hydrogen bonds with H1, and unique covalent bonds with molybdenum atoms underneath, O1 of 1 ML oxide film possesses relatively small negative charge. The magnesium loses electrons to oxygen very easily and the adsorption interaction of molecular water leads to very limited impact on the charging state of surface magnesium. The Mg1 charges of 1 ML - 5 ML systems show little differences. The Mg1 of 1 ML system exhibits smaller positive charge, due to the lack of perpendicular O-Mg ionic bond. Generally, the charge transfer is largest for the 1 ML system, where magnesia film and molybdenum carry charge values + 1.954 e and -1.879 e, respectively.

The atomic structure of water molecule adsorbed on thin magnesia (001) film for molecular adsorption (state A), heterolytic dissociative adsorption (states D1, D2 and D3) and homolytic dissociative adsorption (D4 state) are depicted in Figure 2. The molecular adsorption energies at A state, the heterolytic dissociative adsorption energies at D1 and D2 states, the homolytic dissociative adsorption energies at D3 state are obtained and presented in Table 3. The even-layer oxide films exhibit stronger adsorption interaction toward water molecule with more negative adsorption energies, at A and D1 states. For dissociative adsorption states D2, D3 and D4, the water adsorption on monolayer oxide film exhibits most negative dissociation adsorption energies. The homolytic dissociative adsorption energy of water on monolayer oxide film is calculated to be as large as -1.192 eV. The homolytic dissociative adsorption energy decreases promptly with increase of the oxide thickness (-0.542 eV for 2 ML oxide film), even to positive adsorption energy (0.181 eV, 0.831 eV and 1.509 eV for 3 ML – 5 ML oxide films), indicating the nanoscale (such as thickness) of oxide film plays a crucial role in water dissociation on oxide surface.

For homolytic dissociative structures, the net charges of dissociative water, H1, surface Mg1, Mg2, Mg3 do not exhibit obvious even-odd varying pattern. The charge of dissociative water on 2 ML MgO (001) presents most negative value. The charge H1 is determined to be negative, indicating the formation of hydride ion and the electron accumulating behavior of hydrogen on magnesia-molybdenum hybrid surface. Mg3 carries more positive charge, due to its chemical connection with surface hydroxyl ($O_w$H). The charges of Mg1 and Mg2 are lower than Mg3, because of the larger covalent component of H-Mg bonding than that of O-Mg bonding. On the 1 ML MgO(001)/Mo(001), charge difference between Mg3 and Mg1 is very small, because the Mg3 is two coordinate on 1 ML MgO(001)/Mo(001), and three coordinate on 2 ML – 5 ML MgO(001)/Mo(001) surfaces. The geometric parameters for homolytic dissociative adsorption of water on 1 ML – 5 ML ultrathin oxide films deposited on molybdenum are listed in Table 4. Analogous to the adsorption state, the surface rumpling exhibit even-odd parity effect, in a manner that the surface rumpling for even-layer oxide films are larger than that of neighboring odd-layer oxide films (as shown in Table 4 and Figure 3). However, compared with the rumpling values on pristine magnesia, the surface rumpling on metal-supported magnesia are substantially larger, which are 1.231 Å, 1.088 Å, 1.116 Å, 1.087 Å, and 1.098 Å for 1 ML – 5 ML oxide films respectively. $O_w$-Mg bonding distance are largest for 1 ML oxide film. The optimized H1-Mg1 distances are in the range 1.844 Å ~ 1.860 Å and the H1-Mg2 distances are calculated to be in the range 1.852 Å ~ 1.883 Å, indicating the bonding interaction of H1 with surface magnesium. The sum of H1-Mg1 and H2-Mg2 distances are 3.712 Å, 3.716 Å, 3.719 Å, 3.731 Å and 3.736 Å, indicating the bonding strength are weaker for homolytic dissociative adsorption of water on thick oxide films. The bonding parameters and coordination numbers are more severely changed after the homolytic dissociative adsorption of water on 1 ML – 5ML ultrathin magnesia films deposited on molybdenum (as shown in Table 5). The magnesium at the reaction site are only two coordinate for Mg1 and Mg2. The Mg3 is three coordinate on 2 ML – 5 ML oxide films, and two coordinate on 1 ML oxide film.

**Table 2. Bader charge populations (in electron) of $O_w$, H1, H2, O1, Mg1 atoms, the magnesia films and molybdenum substrates for molecular water adsorption on 1 ML ~ 5 ML ultrathin oxide films deposited on molybdenum, at DFT-D3 theoretical**

level.

| Film thickness | $O_w$ | H1 | H2 | $H_2O$ | O1 | Mg1 | MgO | Mo |
|---|---|---|---|---|---|---|---|---|
| 1 ML | -1.278 | +0.616 | +0.586 | -0.076 | -1.493 | +1.647 | +1.954 | -1.879 |
| 2 ML | -1.286 | +0.606 | +0.575 | -0.105 | -1.588 | +1.669 | +1.473 | -1.368 |
| 3 ML | -1.283 | +0.608 | +0.580 | -0.095 | -1.598 | +1.669 | +1.493 | -1.398 |
| 4 ML | -1.378 | +0.562 | +0.528 | -0.288 | -1.466 | +1.665 | +1.762 | -1.474 |
| 5 ML | -1.283 | +0.604 | +0.583 | -0.096 | -1.600 | +1.670 | +1.484 | -1.388 |

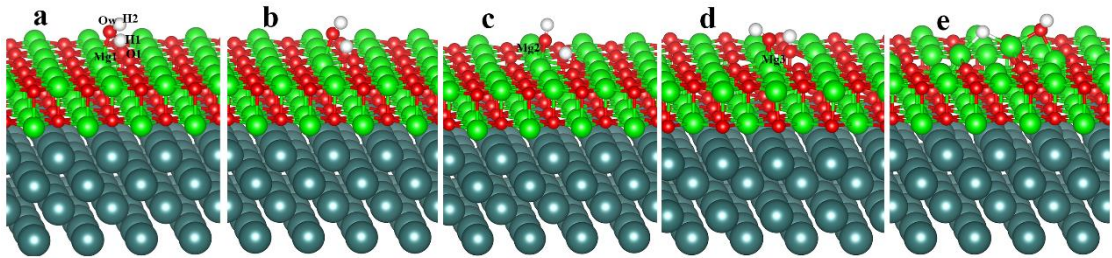

**Figure 2.** The atomic structure of water molecule adsorbed on thin magnesia (001) film. (a) Molecular adsorption, A structure. (b) Heterolytic dissociative adsorption, D1 structure. (c) Heterolytic dissociative adsorption, D2 structure. (d) Heterolytic dissociative adsorption, D3 structure. (e) Homolytic dissociative adsorption, D4 structure.

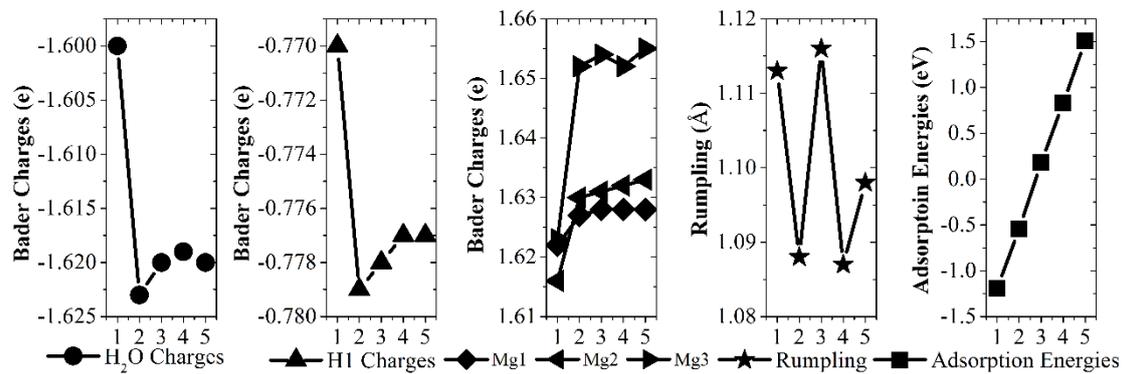

**Figure 3.** $H_2O$ charges, H1 charges, surface Mg charges, surface rumpling, and adsorption energies dependent on film thickness (1 – 5 ML) for homolytically dissociative adsorption of water on molybdenum supported magnesia (001).

**Table 3.** The molecular adsorption energies ($E_{ad}$, in eV) of A state, dissociative

adsorption energies of heterolytic dissociative states D1, D2, D3 and homolytic dissociative D4 states.

| Film thickness | $E_{ad}$(A) | $E_{ad}$(D1) | $E_{ad}$(D2) | $E_{ad}$(D3) | $E_{ad}$(D4) |
|---|---|---|---|---|---|
| 1 ML | -0.805 | —— | -0.947 | -0.298 | -1.192 |
| 2 ML | -0.853 | -0.865 | -0.836 | -0.088 | -0.542 |
| 3 ML | -0.826 | -0.795 | -0.722 | 0.121 | 0.181 |
| 4 ML | -0.854 | -0.833 | -0.757 | 0.080 | 0.831 |
| 5 ML | -0.832 | -0.808 | -0.728 | 0.114 | 1.509 |

**Table 4. The geometric parameters and surface rumpling (in angstrom) for homolytic dissociative adsorption of water on 1 ML ~ 5 ML ultrathin oxide films deposited on molybdenum, at DFT-D3 theoretical level.**

| Film thickness | H1-Mg1 | H1-Mg2 | Ow-Mg1 | Ow-Mg3 | $\Delta z^a$ |
|---|---|---|---|---|---|
| 1 ML | 1.860 | 1.852 | 2.006 | 2.053 | 1.231 |
| 2 ML | 1.844 | 1.872 | 1.968 | 1.984 | 1.088 |
| 3 ML | 1.848 | 1.871 | 1.965 | 1.985 | 1.116 |
| 4 ML | 1.851 | 1.880 | 1.971 | 1.987 | 1.087 |
| 5 ML | 1.853 | 1.883 | 1.970 | 1.991 | 1.098 |

$^a$ The surface rumpling are defined as the projected distance of the top atom and the bottom atom of the first layer, in $z$ direction.

When the hydroxyl of molecularly adsorbing water is broken, the surface hydroxyl O1H1 and O$_w$H are obtained and first heterolytic dissociative state D1 take place, by crossing slight activation barriers 0.006 eV, 0.031 eV, 0.023 eV and 0.024 eV for 2 ML – 5 ML oxide films, respectively. The D1 state on 2 ML oxide film shows lowest relative energy (i.e., largest adsorption energy) and smallest activation barrier, as shown in Figure 4. Compared with the neighboring odd-layer magnesia films and bulk magnesia, the even-layer ultrathin magnesia films facilitate the heterolytic dissociation in A state. The D2 configuration on 2 ML – 5 ML oxide films can be obtained by translating the hydroxyl group to bind with two surface magnesium, with slight energy elevation. The energy barrier for transformation reaction from D1 to D2 over metal-supported magnesia (001) is calculated to be 0.046 eV, 0.080 eV, 0.083 eV and 0.085 eV for 2 ML – 5 ML oxide films, respectively (as shown in Figure 5). For the monolayer oxide film, the D2 configuration can be formed directly from the exothermic transformation reaction from A state to D2 state, with energy release of -0.142 eV and small activation barrier of 0.011 eV (as shown in Figure 6). The D3 structure can be formed either directly from molecular adsorption state (structure A), or from another heterolytic dissociative adsorption state D2, as shown in Figures 7 and 8. The dissociation pathway from A state to D3 state shows activation energy barriers of 0.740 eV, 0.854 eV, 1.040 eV, 1.022 eV and 1.037 eV for 1 ML – 5 ML oxide films, respectively (Figure 7 and Table 6). The transformation reaction between heterolytic dissociative state D2 and D3 present energy barriers 0.885 eV, 0.842 eV, 0.942 eV, 0.952 eV and 0.900 eV, respectively for transformation processes on 1 ML – 5 ML oxide films. Generally considering the two pathways, the 1 - 2 ML ultrathin oxide films are more favorable for dissociating water to form D3 state. The homolytic dissociative adsorption structure D4 could be obtained by transformation reaction from heterolytic dissociative adsorption structure D3 (as shown in Figure 9), presenting activation barriers 0.733 eV, 1.103 eV, 1.306 eV, 1.571 eV, and 1.849 eV. After dissociation reaction, the hydrogen of surface hydroxyl O1H1 translate to bind chemically with two surface magnesium. It is worth to note that, compared with the heterolytic dissociative states, the newly formed homolytic dissociative adsorption state on 1 ML MgO(001)/Mo(001) possesses significantly larger adsorption energy (-1.192 eV) and lower activation energy (0.733 eV), demonstrating the hydride species are stabilized on ultrathin magnesia.

**Table 5. The bonding parameters and coordination numbers (CN) of surface magnesium at the dissociative reaction sites for homolytic dissociative adsorption of water on 1 ML ~ 5 ML ultrathin oxide films deposited on molybdenum, at DFT-D3 theoretical level.**

| Film thickness | Mg1-O [a] | Mg2-O [a] | Mg3-O [a] | CN(Mg1, Mg2, Mg3) [b] |
|---|---|---|---|---|
| 1 ML | **2.138, 2.328,** 2.980, 3.025 | **2.083, 2.097,** 2.704, 2.806 | **2.069, 2.101,** 2.574, 2.781 | 2, 2, 2 |
| 2 ML | **2.050, 2.061,** *3.294,* 3.443, 3.458 | **1.935, 1.936,** *2.882, 3.189,* 3.446, 3.453 | **2.000, 2.016, 2.201,** 2.668, 2.744 | 2, 2, 3 |
| 3 ML | **2.057, 2.063,** *3.386,* 3.431, 3.435 | **1.936, 1.937,** *2.992, 3.315,* 3.436, 3.438 | **2.002, 2.024, 2.255,** 2.623, 2.723 | 2, 2, 3 |
| 4 ML | **2.057, 2.063,** *3.327,* 3.416, 3.428 | **1.936, 1.936,** *2.936, 3.287,* 3.404, 3.418 | **2.007, 2.026, 2.250,** 2.629, 2.720 | 2, 2, 3 |
| 5 ML | **2.054, 2.063,** *3.341,* 3.414, 3.417 | **1.937, 1.937,** *2.934, 3.290,* 3.408, 3.411 | **2.005, 2.027, 2.254,** 2.617, 2.717 | 2, 2, 3 |

[a] The bonds between magnesia of first layer and oxygen of second layer are given in italic fonts. The bond distances smaller than 2.500 are emphasized in bold fonts.

[b] Coordination numbers listed here do not take into account the bonds with distances larger than 2.5 Å.

**Table 6.** Activation energy barriers ($E_a$, in eV) of dissociation and transformation reaction processes A → D3, D2 → D3, and D3 → D4.

| Film thickness | $E_a$(A → D3) | $E_a$(D2 → D3) | $E_a$(D3 → D4) |
|---|---|---|---|
| 1 ML | 0.740 | 0.885 | 0.733 |
| 2 ML | 0.854 | 0.842 | 1.103 |
| 3 ML | 1.040 | 0.942 | 1.306 |
| 4 ML | 1.022 | 0.925 | 1.571 |
| 5 ML | 1.037 | 0.900 | 1.849 |

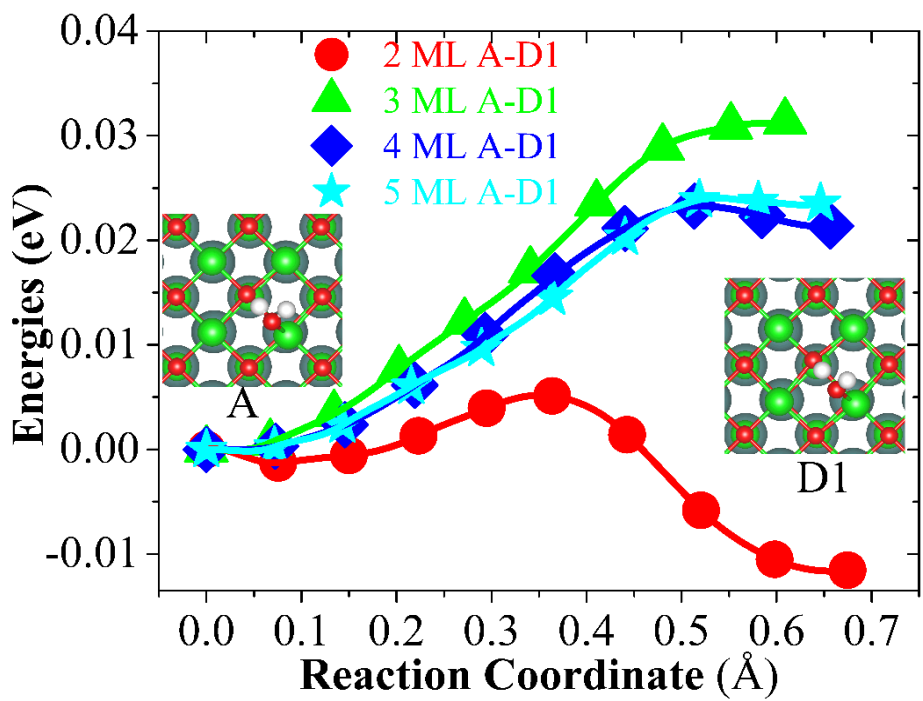

**Figure 4.** The reaction pathways for dissociative state D1 of water on 2 ML ~ 5 ML ultrathin oxide films deposited on molybdenum, at DFT-D3 theoretical level.

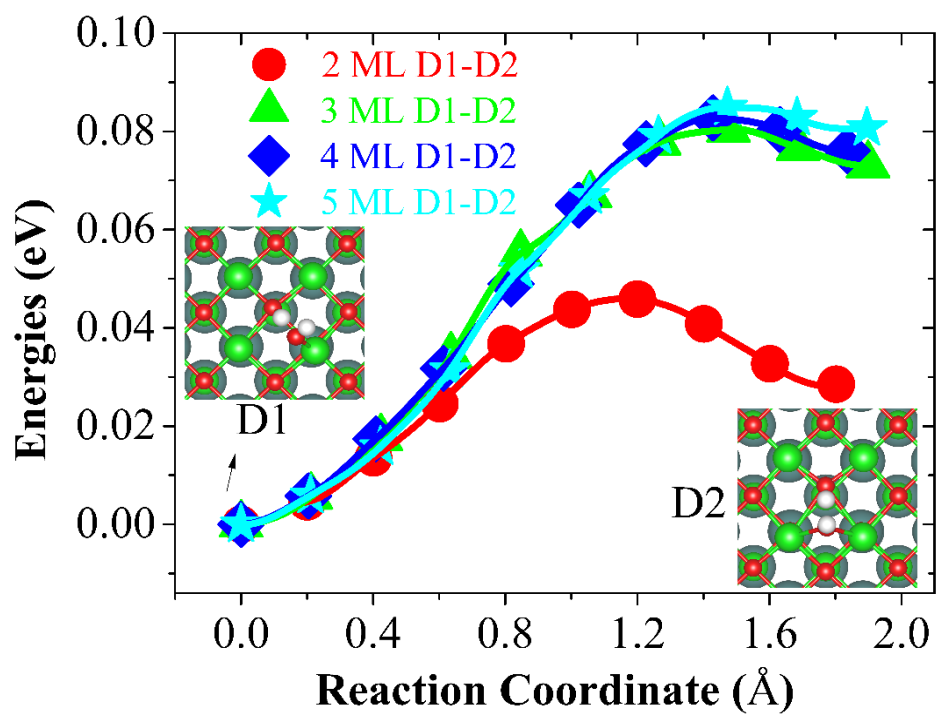

**Figure 5.** The reaction pathways for transformation from dissociative state D1 to D2 for water adsorption on 2 ML ~ 5 ML ultrathin oxide films deposited on molybdenum, at DFT-D3 theoretical level.

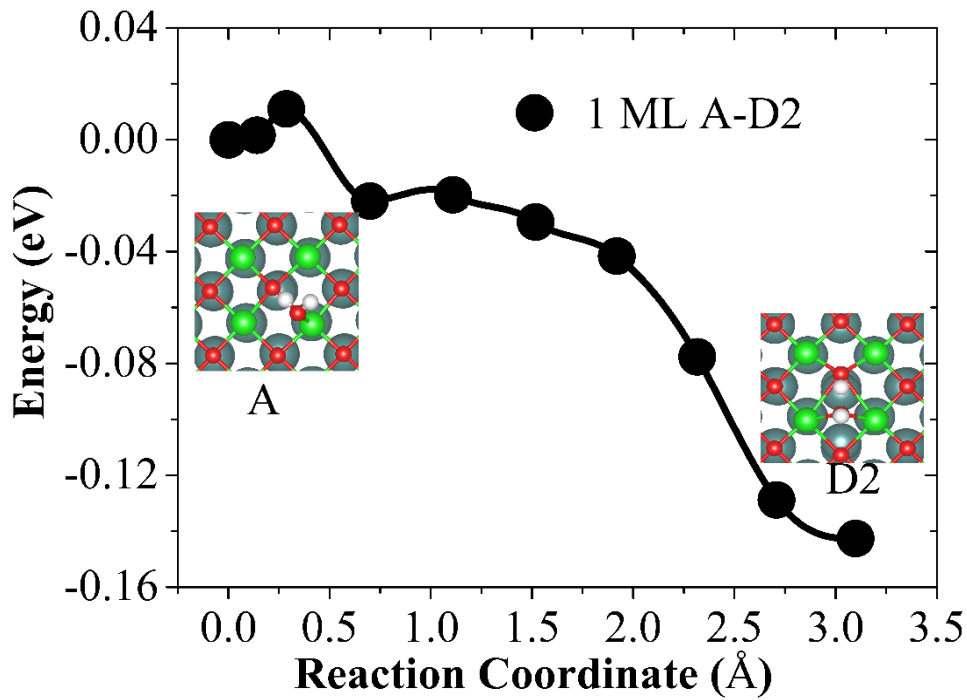

**Figure 6.** The reaction pathway for dissociative state D2 of water on 1 ML ultrathin oxide films deposited on molybdenum, at DFT-D3 theoretical level.

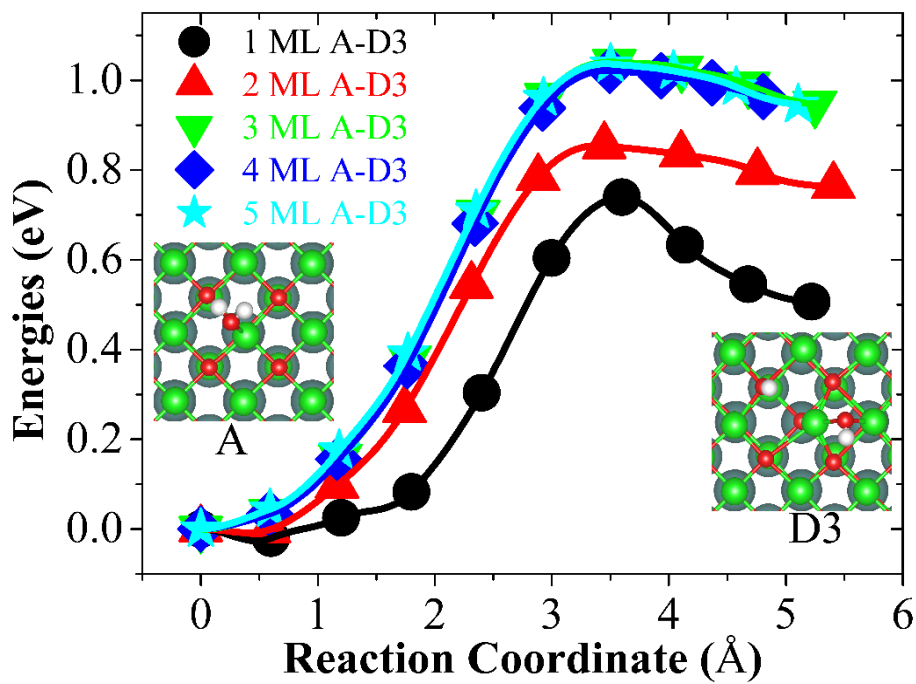

**Figure 7.** The reaction pathways for dissociative state D3 of water on 1 ML ultrathin oxide films deposited on molybdenum, at DFT-D3 theoretical level.

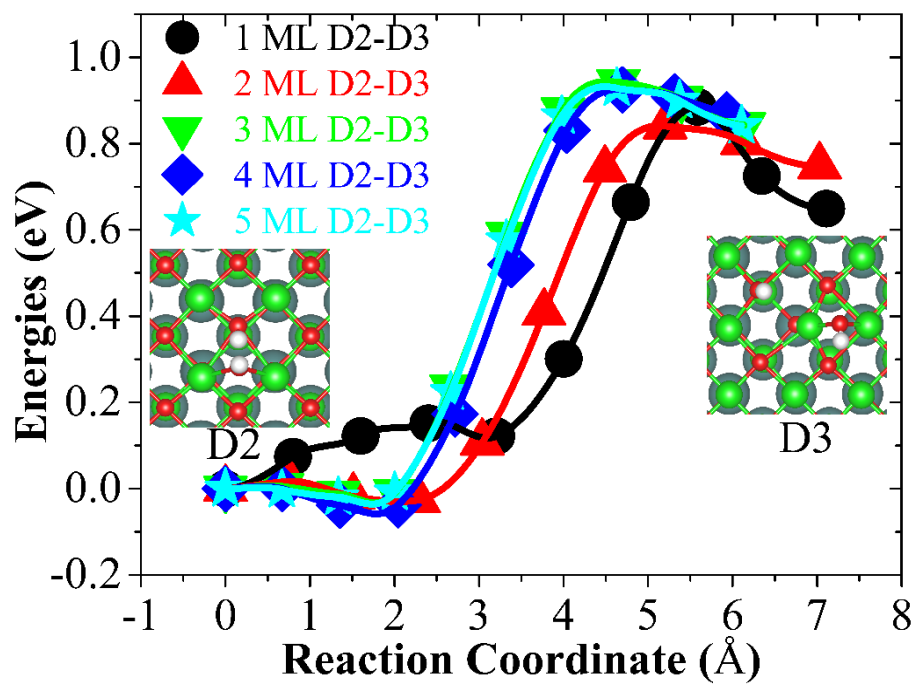

**Figure 8.** The reaction pathways for transformation from dissociative state D2 to D3 for water adsorption on 1 ML ~ 5 ML ultrathin oxide films deposited on molybdenum, at DFT-D3 theoretical level.

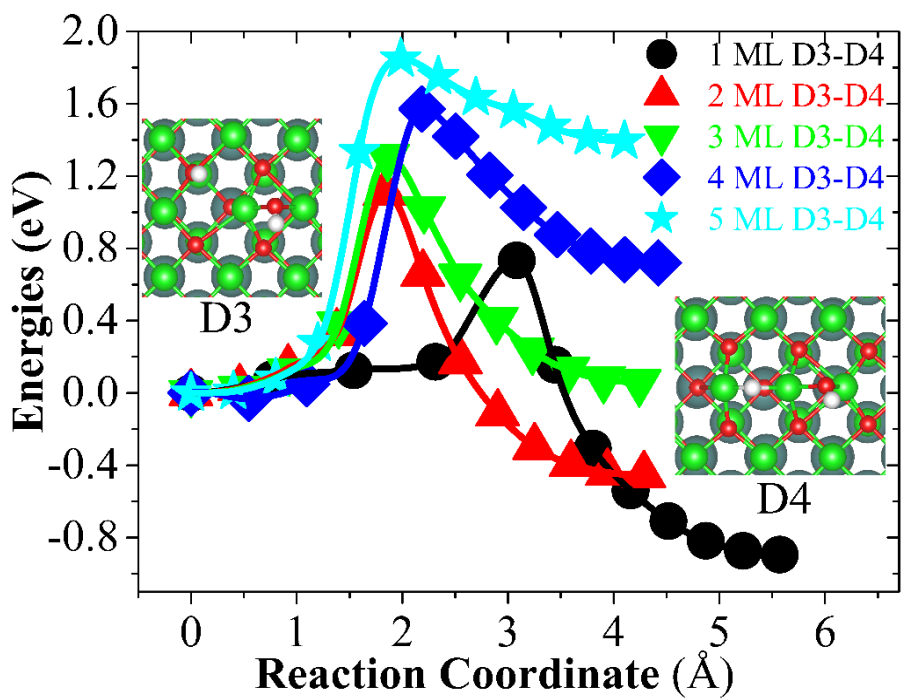

**Figure 9.** The reaction pathways for homolytically dissociative state D4 for water adsorption on 1 ML ~ 5 ML ultrathin oxide films deposited on molybdenum, at DFT-D3 theoretical level.

**3.3 The electronic properties for homolytic dissociative adsorption of water on MgO/Mo**

For the homolytic dissociative adsorption of water, we calculated its Bader charge populations of $O_wH$, H1, $H_2O$, O1, Mg1, the magnesia films and molybdenum substrates to interpret the charge transfer mechanism in the hybrid structure and between surface and adsorbates. $O_wH2$ charges are in the range of -0.830 e ~ -0.844 e, indicating the direct production of a hydroxyl group, which is similar to heterolytic dissociation of water.[35] In particular, the charges of H1 are negative with value in the range of -0.770 e ~ -0.779 e, confirming the definite formation of surface hydride, which are markedly differently from all the reported heterolytic dissociative adsorption structures. The total charges of dissociated water are in the range of -1.600 e ~ -1.623 e. The slightly smaller charge of dissociated water on one monolayer oxide film is due to the large charge transfer between the surface atoms of magnesia (001) to underneath molybdenum substrate, which to some extent affects the ability of electron accumulation of dissociated water. The surface magnesium Mg1, Mg2, Mg3 possess least charge value at 1 ML oxide surface. This result can be ascribed to the large charge transfer from magnesia to molybdenum substrate and the sufficient bonding interaction between oxygen and molybdenum. Consequently, the sufficiency and saturation of ionic bonding for surface magnesium are lowered for monolayer oxide film. Compared with Mg1 and Mg2, at the reaction site, the charge of Mg3 is largest, for Mg3 is connected chemically to surface hydroxyl, and Mg3 possesses higher coordination number. The molybdenum substrates in all hybrid surfaces show negative charges, and the molybdenum substrates under odd-layer oxide films show more negative charges than that under even-layer oxide films. The oxide film exhibit large positive charges, and the charge values show obvious even-odd alteration effect (+2.342 e, +2.109 e, +2.200 e, +2.054 e and +2.270 e for 1 ML – 5 ML magnesia films, respectively). It can be speculated that the odd-layer oxide films show stronger reducibility than even-layer oxide films, and are more severely oxidized by adsorbates and metal substrate with high electron affinity.

The local density of states of H1, H2 and $O_wH2$ are analyzed to reveal the electronic structure of homolytic dissociative water on ultrathin magnesia films deposited on molybdenum (Figure 10). Compared with other adsorbates, the electronic states of H1 generally locate at lower energy scopes, for the bonding orbital of H1 is mainly distributed in 1s with low energy level. The states of H1 are closer to Fermi level, with increasing film thickness. This result implies the thick oxide films is unfavorable for stabilizing the produced hydride species. The electronic states of H2 and $O_wH2$ exhibit similar tendency and same peak positions at -3.4 eV, -2.8 eV and -2.2 eV, because of the strong covalent bonding between H2 and $O_w$. The highest occupied states of produced surface hydroxyl are shifted to energy levels near Fermi level, with the increase of film thickness. While the hydroxyl $O_wH$ on thinner oxide films are stabilized to lower energy levels. States of Mg1, Mg2, Mg3 experience complex fluctuation, indicating the multiple bonding with surrounding negatively charged species (Figure 11). The state intensity of surface magnesium at reaction site is very weak, which implies the majority of electrons of surface magnesium are acquired by neighboring oxygen, hydride, and hydroxyl species. At energy levels around -3.3 eV, -2.8 eV and -2.2 eV, the states of Mg3 in 1 ML-3 ML oxide films are significantly weaker than Mg1 and Mg2, suggesting that Mg3 are oxidized more adequately, and the electrons of Mg3 at high energy level are grasped readily by the coordinated oxygen and dissociated adsorbates. Compared with Mg2, the state peaks of Mg1 in 1 ML oxide film, at energy level around -3.3 eV are enhanced. In addition, the states of Mg1 in 2 ML and 3 ML oxide films are split at energy levels around -2.8 eV and -2.2 eV, respectively. The results can be ascribed to the multiple coordination with produced hydride and hydroxyl species. The electronic states for interfacial molybdenum atoms and interfacial oxygen atoms show broad overlap between Mo-$4d_z^2$ and O-$2p_z$ orbitals at energy levels between -2.5 eV and -6.6 eV (as shown in Figure 12), indicating the effective hybridization and formation of chemical bonding interaction. For the 1 ML oxide film, large amount of electrons occupy states in the range of -4.5 eV to -5.5 eV, while the intensity between -6 eV and -6.6 eV is very weak. This result is obviously different from the states distribution of 2 ML and 3 ML oxide films, indicating the unique coordination environment and unsaturated character of interfacial oxygen in 1 ML oxide film.

The differential charge density contour for homolytic dissociative adsorption of water on monolayer magnesia deposited on molybdenum is shown in Figure 13. Very different from the heterolytic dissociation structures, the homolytic dissociation products all show electron accumulation effects, which agrees well with the Bader charge population. Due to the electron-abstracting property of adsorbate species, the top of oxygen around reaction site presents cyan color, indicating the electron depletion at this area. All the interfacial oxygen form chemical bonds with interfacial molybdenum, and the shared electron pairs shift towards oxygen because of the larger electron affinity of oxygen. The obvious electron accumulation with columnar shape occurs at the interface area, indicating the effective charge transfer and binding interaction between magnesia and metal substrate, even under the circumstance of strong oxidizing adsorbates (surface hydroxyl) and severe surface reconstruction. Although the electron pairs between oxygen and molybdenum are shifted toward interfacial oxygen, the molybdenum inner bulk exhibit electron accumulation effects, which is responsible for the general negative charge of metal substrate.

The electron localization function (ELF) is analyzed to provide appropriate and vivid description of the complex chemical bonds in the oxide/metal hybrid structure and between the hybrid structure and adsorbates (Figure 14). From the ELF contour, we could clearly see the obvious feature of electron localization for the produced hydride (H1) and hydroxyl ($O_wH2$) on the magnesia surface, indicating the formation of ionic bonds between adsorbates and metal-supported magnesia. Different from the bonding of split water and magnesia, the bulk molybdenum exhibit strong electron delocalization due to its metallic nature. As depicted in the highest occupied orbital (Figure 15), for the homolytic dissociation of water on MgO(001)/Mo, the electrons with highest energy level mainly distribute in molybdenum substrate. Due to the bonding interaction between surface oxygen and molybdenum, small pieces of highest occupied orbital extend to the interfacial oxygen. The bonding orbital of H-Mg, with *s* orbital character, is depicted in Figure 15b. Although the interaction between hydroxyl of dissociated water and magnesia is studied extensively, bonding interaction between hydride of dissociated water with perfect magnesia is revealed for the first time in this work.

**Table 7. Bader charge populations (in electron) of $O_wH$, H1, $H_2O$, O1, Mg1, the magnesia films and molybdenum substrates for homolytic dissociative adsorption of water on 1 ML ~ 5 ML ultrathin oxide films deposited on molybdenum, at DFT-D3 theoretical level.**

| Film thickness | $O_wH2$ | H1 | $H_2O$ | Mg1 | Mg2 | Mg3 | MgO | Mo |
|---|---|---|---|---|---|---|---|---|
| 1 ML | -0.830 | -0.770 | -1.600 | +1.622 | +1.616 | +1.623 | +2.342 | -0.741 |
| 2 ML | -0.844 | -0.779 | -1.623 | +1.627 | +1.630 | +1.652 | +2.109 | -0.486 |
| 3 ML | -0.842 | -0.778 | -1.620 | +1.628 | +1.631 | +1.654 | +2.200 | -0.581 |
| 4 ML | -0.842 | -0.777 | -1.619 | +1.628 | +1.632 | +1.652 | +2.054 | -0.436 |
| 5 ML | -0.843 | -0.777 | -1.620 | +1.628 | +1.633 | +1.655 | +2.270 | -0.650 |

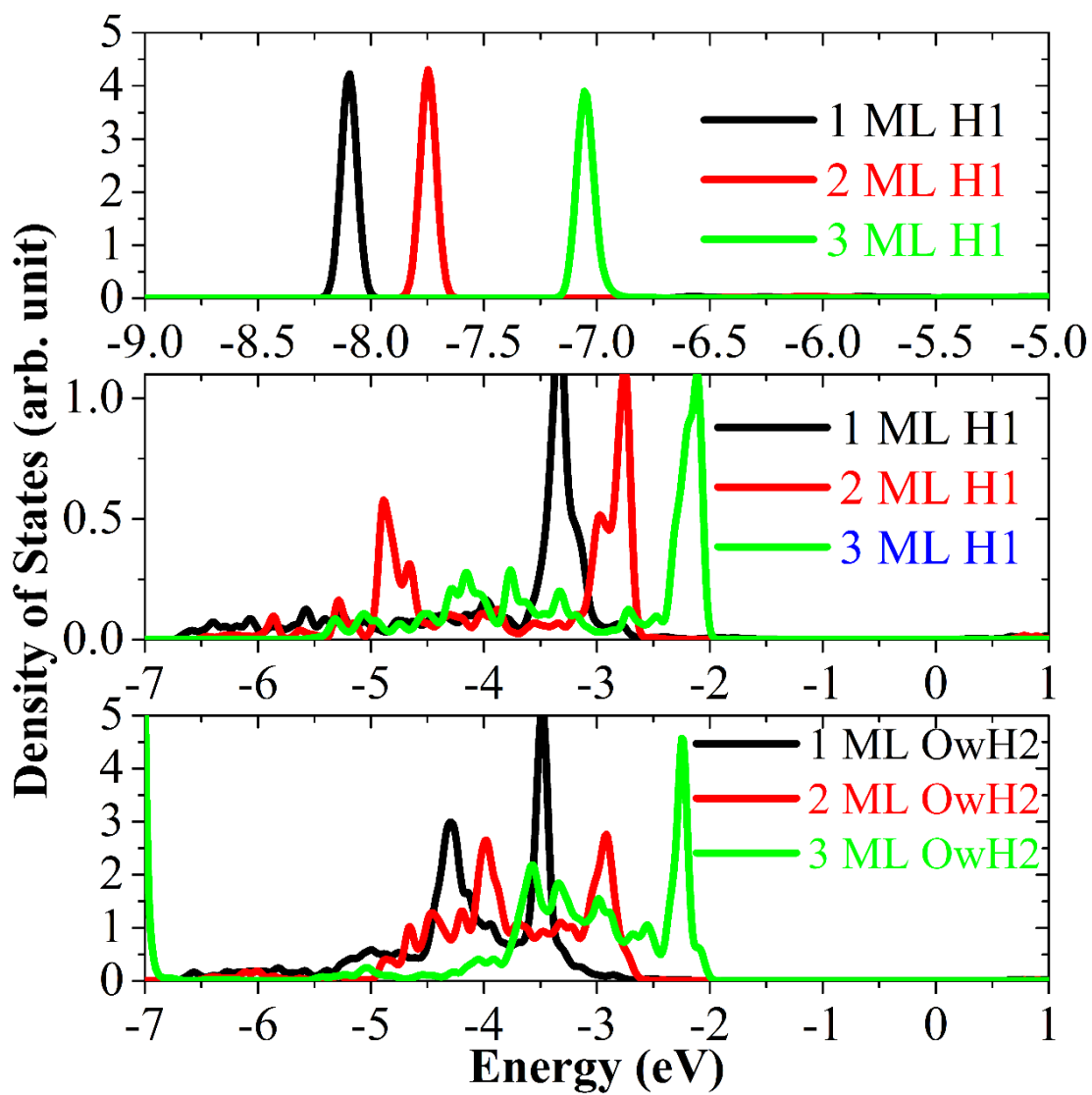

**Figure 10.** Local density of states of H1, H2 and OwH2 for homolytically dissociative water on 1 – 3 monolayer magnesia deposited on molybdenum. The energies are relative to Fermi energy level.

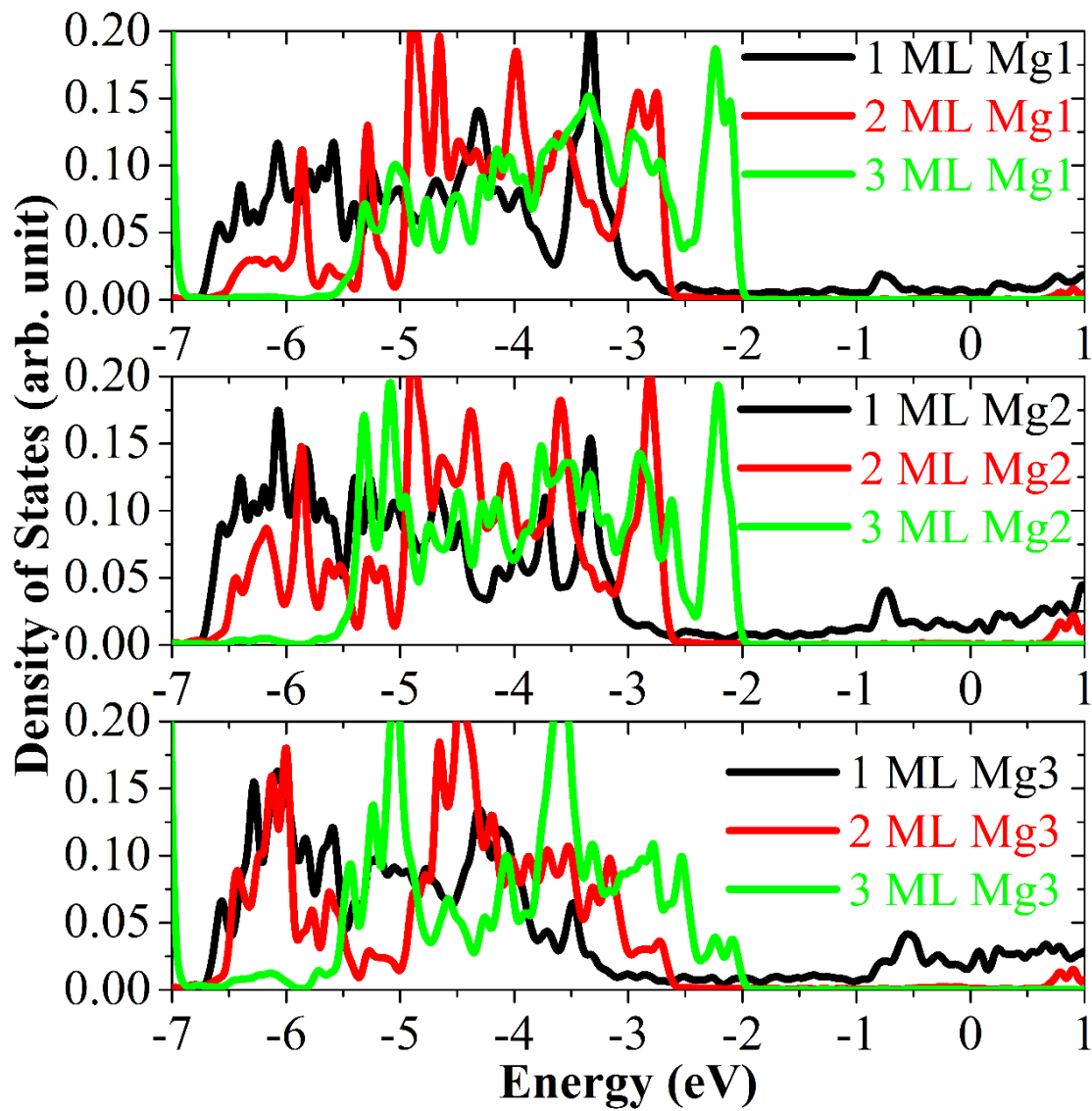

**Figure 11.** Local density of states of Mg1, Mg2 and Mg3 for homolytically dissociative water on 1-3 monolayer magnesia deposited on molybdenum. The energies are relative to Fermi energy level.

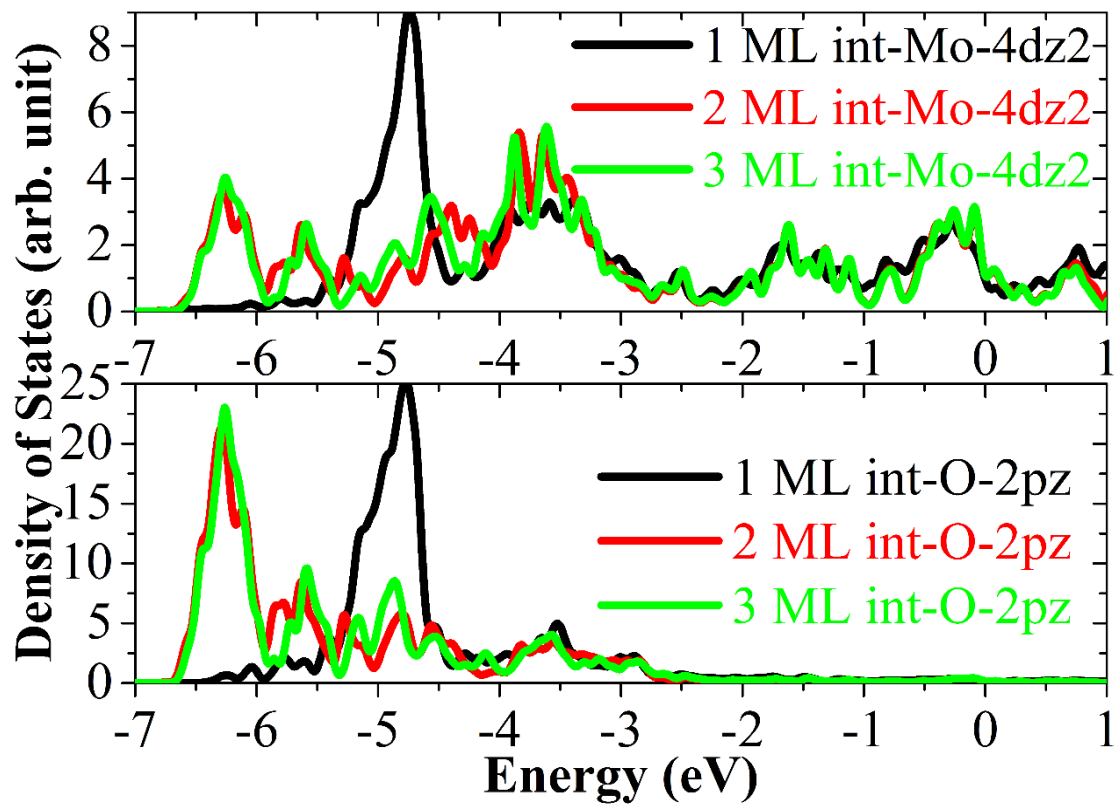

**Figure 12.** Projected density of states of interfacial molybdenum atoms and interfacial oxygen atoms for homolytically dissociative water adsorbed on 1-3 monolayer magnesia deposited on molybdenum. The energies are relative to Fermi energy level.

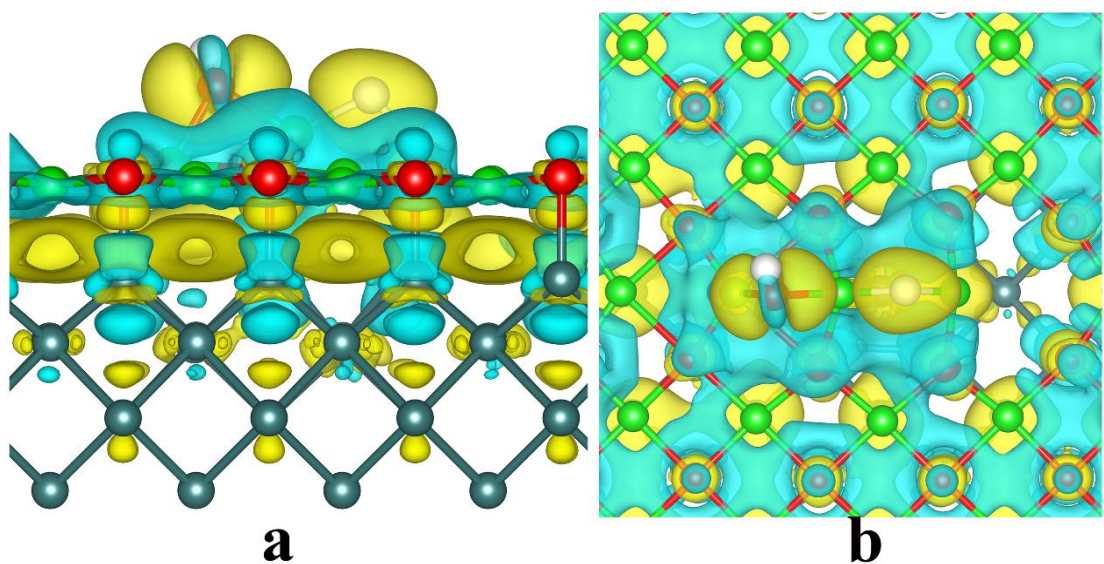

**Figure 13.** Differential charge density contour for water adsorbing on monolayer magnesia deposited on molybdenum, with homolytically dissociative configuration. (a) Side view, (b) Top view. The differential charge densities is defined as $\Delta\rho = \rho(\text{Total}) -$

ρ(H) - ρ(OH) - ρ(MgO) - ρ(Mo). The isosurface value for the differential charge densities is 0.002 e Bohr$^{-3}$. The yellow and cyan colors stand for electron accumulation and electron depletion, respectively.

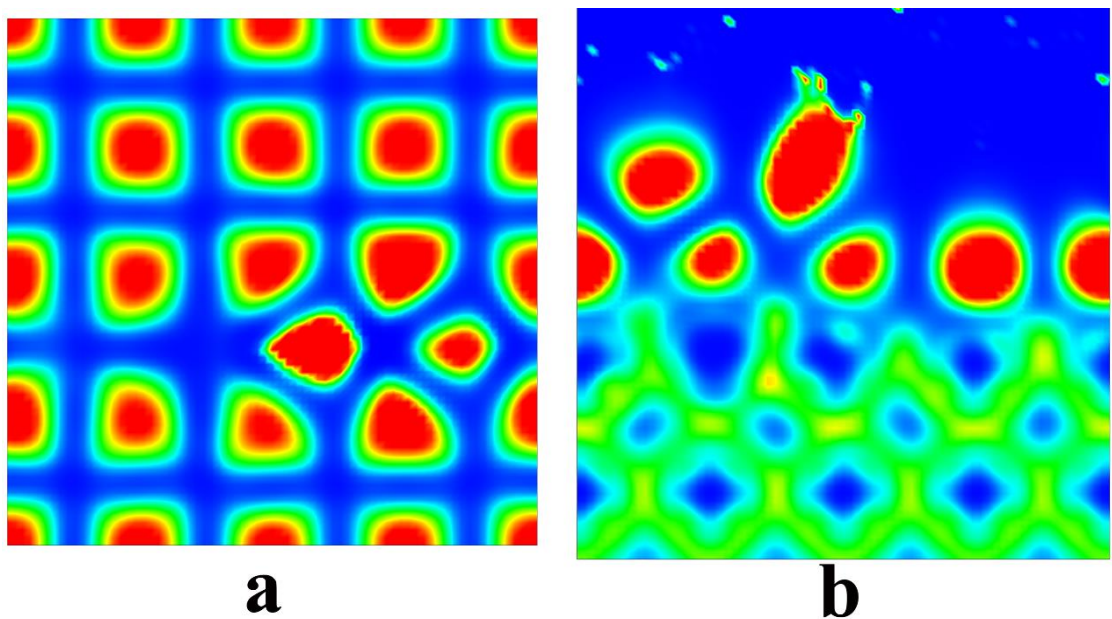

**Figure 14.** Electron localization function for homolytically dissociative water adsorbed on monolayer magnesia deposited on molybdenum. (a) Top view, (b) Side view. The plots of electron localization function are under the same saturation levels. The blue and red regions stand for electron delocalization and electron localization, respectively.

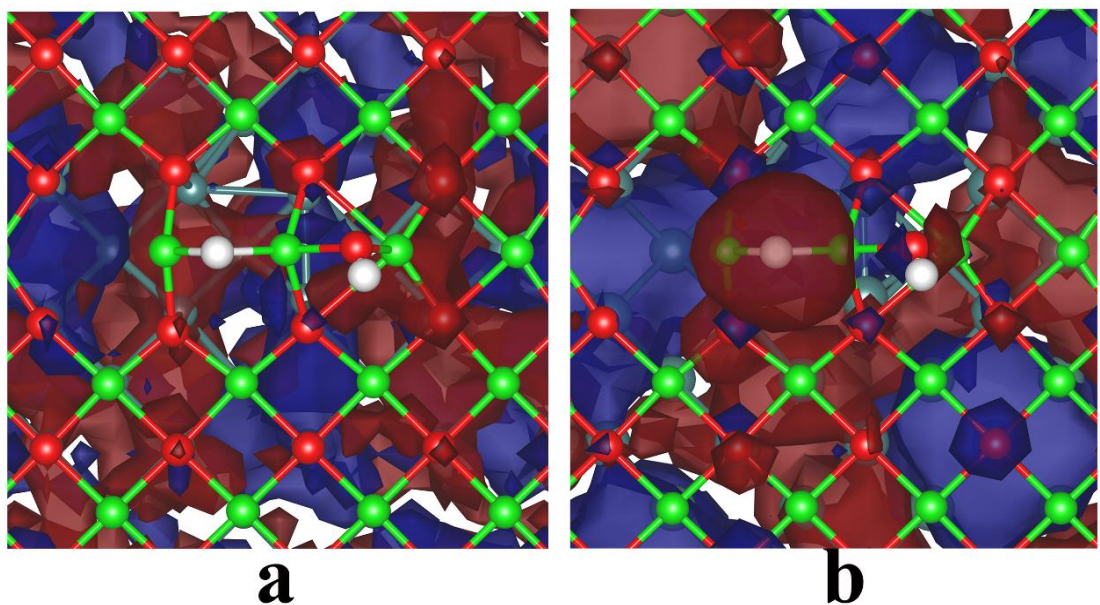

**Figure 15.** (a) Highest occupied molecular orbital for homolytically dissociative water adsorbed on monolayer magnesia deposited on molybdenum. (b) The orbital with largest orbital coefficient for H-Mg bonding interaction for homolytically dissociative water adsorbed on monolayer magnesia deposited on molybdenum. The isosurface values for orbitals are set to be 0.005 e Å$^{-3}$.

## 4. Conclusion

In summary, by utilizing periodic density functional theory calculations with van der Waals corrections, we report for the first time the energetically favorable homolytic dissociative adsorption behavior of water over magnesia (001) films deposited on metal substrate. The water adsorption on pristine magnesia (001) is quite weak and the heterolytic dissociation is the only fragmentation pathway, which is highly endothermic by 1.157 eV with large activation barrier of 1.167 eV. The binding strength for the molecular and dissociative adsorption configurations of water on MgO(001)/Mo(001) are significantly larger than corresponding configurations on bare magnesia slab. The homolytic dissociative adsorption energy of water on monolayer oxide is calculated to be as large as -1.192 eV, which is even larger than all the heterolytic dissociative adsorption states. With the increase of the oxide thickness, the homolytic dissociative adsorption energy decreases promptly (-0.542 eV for 2 ML oxide film), even to positive adsorption energy (0.181 eV, 0.831 eV and 1.509 eV for 3 ML – 5 ML oxide films), indicating the nanoscale (such as thickness) of oxide film play a crucial role in homolytic splitting of water on oxide surface. When the hydroxyl of molecularly adsorbing water is broken, the surface hydroxyl O1H1 and $O_wH$ are obtained and first heterolytic dissociative state D1 take place, by crossing slight activation barriers 0.006 eV, 0.031 eV, 0.023 eV and 0.024 eV for 2 ML – 5 ML oxide films, respectively. The energy barrier for transformation reaction from D1 to D2 over metal-supported magnesia (001) is calculated to be 0.046 eV, 0.080 eV, 0.083 eV and 0.085 eV for 2 ML – 5 ML oxide films, respectively. For the monolayer oxide film, the D2 configuration can be formed directly from the exothermic transformation reaction from A state to D2 state, with energy release of -0.142 eV and small activation barrier of 0.011 eV. The D3 structure can be formed either directly from molecular adsorption state (structure A), or from another heterolytic dissociative adsorption state D2. The dissociation pathway from A state to D3 state shows activation energy barriers of 0.740 eV, 0.854 eV, 1.040 eV, 1.022 eV and 1.037 eV for 1 ML – 5 ML oxide films, respectively. The transformation reaction between heterolytic dissociative state D2 and D3 present energy barriers 0.885 eV, 0.842 eV, 0.942 eV, 0.952 eV and 0.900 eV, respectively for transformation processes on 1 ML – 5 ML oxide films. Generally considering the two pathways, the 1 - 2 ML ultrathin oxide films are more favorable for dissociating water to form D3 state. The homolytic dissociative adsorption structure D4 could be obtained by transformation reaction from heterolytic dissociative adsorption structure D3, presenting activation barriers 0.733 eV, 1.103 eV, 1.306 eV, 1.571 eV, and 1.849 eV. The surface rumpling for homolytic splitting of water on metal-supported ultrathin magnesia (001) is substantially larger than that on bare magnesia (001) and exhibits even-odd parity effect, in a manner that the surface rumpling for even-layer oxide film is larger than that of neighboring odd-layer oxide film. The bonding parameters and coordination numbers are more severely changed after the homolytic dissociative adsorption of water on 1 ML – 5ML ultrathin magnesia films deposited on molybdenum. Bader charge population and differential charge density contours are analyzed to interpret charge transfer mechanism in the hybrid structure and between surface and adsorbates. The definite formation of surface hydride with negative charges -0.770 e ~ -0.779 e and the total charges of dissociated water -1.600 e ~ -1.623 e are markedly different from the heterolytic dissociative adsorption state and the molecular adsorption state. The electronic states of dissociative adsorbates are located at lower relative energy ranges, with decreasing film thickness, which implies the ultrathin oxide films (1 ML and 2 ML) are more favorable for stabilizing the produced hydride and hydroxyl species. The bonding characteristics between hydride/hydroxyl of homolytically dissociated water with ultrathin magnesia (001) is uncovered in this work by characterizing the electron localization function and particular occupied orbitals. It is anticipated the result here should provide inspiring clue and versatile strategy for enhancing chemical reactivity and properties of insulating oxide toward homolytic fragmentation of water.

## ACKNOWLEDGMENTS

This work was supported by the NSFC (Grants 21625103, 21571107, and 21421001), Project 111 (Grant B12015), and the SFC of Tianjin (Grant 15JCZDJC37700). The density functional calculations in this research were performed on TianHe-1(A) at National Supercomputer Center in Tianjin.

## REFERENCES

(1) (a) Karim, W.; Spreafico, C.; Kleibert, A.; Gobrecht, J.; VandeVondele, J.; Ekinci, Y.; van Bokhoven, J. A., Catalyst support effects on hydrogen spillover. *Nature* **2017**, *541*, 68-71;   (b) Kuhlenbeck, H.; Shaikhutdinov, S.; Freund, H. J., Well-Ordered Transition Metal Oxide Layers in Model Catalysis - A Series of Case Studies. *Chemical Reviews* **2013**, *113*, 3986-4034;   (c) Nilius, N.; Freund, H. J., Activating Nonreducible Oxides via Doping. *Accounts of Chemical Research* **2015**, *48*, 1532-1539.

(2) (a) Konig, T.; Simon, G. H.; Martinez, U.; Giordano, L.; Pacchioni, G.; Heyde, M.; Freund, H. J., Direct Measurement of the Attractive Interaction Forces on F-0 Color Centers on MgO(001) by Dynamic Force Microscopy. *Acs Nano* **2010**, *4*, 2510-2514;   (b) Puigdollers, A. R.; Schlexer, P.; Tosoni, S.; Pacchioni, G., Increasing Oxide

Reducibility: The Role of Metal/Oxide Interfaces in the Formation of Oxygen Vacancies. *Acs Catalysis* **2017**, *7*, 6493-6513;   (c) Yang, F.; Choi, Y.; Liu, P.; Stacchiola, D.; Hrbek, J.; Rodriguez, J. A., Identification of 5-7 Defects in a Copper Oxide Surface. *Journal of the American Chemical Society* **2011**, *133*, 11474-11477.

(3) (a) Roldan, A.; Ricart, J. M.; Illas, F.; Pacchioni, G., O-2 Activation by Au-5 Clusters Stabilized on Clean and Electron-Rich MgO Stepped Surfaces. *Journal of Physical Chemistry C* **2010**, *114*, 16973-16978;   (b) Lin, X.; Nilius, N.; Sterrer, M.; Koskinen, P.; Hakkinen, H.; Freund, H. J., Characterizing low-coordinated atoms at the periphery of MgO-supported Au islands using scanning tunneling microscopy and electronic structure calculations. *Physical Review B* **2010**, *81*;   (c) Nilius, N.; Kozlov, S. M.; Jerratsch, J. F.; Baron, M.; Shao, X.; Vines, F.; Shaikhutdinov, S.; Neyman, K. M.; Freund, H. J., Formation of One-Dimensional Electronic States along the Step Edges of CeO2(111). *Acs Nano* **2012**, *6*, 1126-1133.

(4) (a) Yoon, B.; Hakkinen, H.; Landman, U.; Worz, A. S.; Antonietti, J. M.; Abbet, S.; Judai, K.; Heiz, U., Charging effects on bonding and catalyzed oxidation of CO on Au-8 clusters on MgO. *Science* **2005**, *307*, 403-407;   (b) Sterrer, M.; Fischbach, E.; Risse, T.; Freund, H. J., Geometric characterization of a singly charged oxygen vacancy on a single-crystalline MgO(001) film by electron paramagnetic resonance spectroscopy. *Physical Review Letters* **2005**, *94*;   (c) Konig, T.; Simon, G. H.; Rust, H. P.; Pacchioni, G.; Heyde, M.; Freund, H. J., Measuring the Charge State of Point Defects on MgO/Ag(001). *Journal of the American Chemical Society* **2009**, *131*, 17544-17545;   (d) Pacchioni, G., Numerical Simulations of Defective Structures: The Nature of Oxygen Vacancy in Non-reducible (MgO, SiO2, ZrO2) and Reducible (TiO2, NiO, WO3) Oxides. In *Defects at Oxide Surfaces*, Jupille, J.; Thornton, G., Eds. Springer International Publishing: Cham, 2015; pp 1-28.

(5) (a) Lin, X., et al., Charge-Mediated Adsorption Behavior of CO on MgO-Supported Au Clusters. *Journal of the American Chemical Society* **2010**, *132*, 7745-7749;   (b) Brown, M. A.; Carrasco, E.; Sterrer, M.; Freund, H. J., Enhanced Stability of Gold Clusters Supported on Hydroxylated MgO(001) Surfaces. *Journal of the American Chemical Society* **2010**, *132*, 4064-+;   (c) Risse, T.; Shaikhutdinov, S.; Nilius, N.; Sterrer, M.; Freund, H. J., Gold supported on thin oxide films: From single atoms to nanoparticles. *Accounts of Chemical Research* **2008**, *41*, 949-956;   (d) Chen, M. S.; Goodman, D. W., The structure of catalytically active gold on titania. *Science* **2004**, *306*, 252-255.

(6) Senanayake, S. D.; Stacchiola, D.; Rodriguez, J. A., Unique Properties of Ceria Nanoparticles Supported on Metals: Novel Inverse Ceria/Copper Catalysts for CO Oxidation and the Water-Gas Shift Reaction. *Accounts of Chemical Research* **2013**, *46*, 1702-1711.

(7) Rodriguez, J. A.; Senanayake, S. D.; Stacchiola, D.; Liu, P.; Hrbek, J., The Activation of Gold and the Water-Gas Shift Reaction: Insights from Studies with Model Catalysts. *Accounts of Chemical Research* **2014**, *47*, 773-782.

(8) Rodriguez, J. A.; Liu, P.; Stacchiola, D. J.; Senanayake, S. D.; White, M. G.; Chen, J. G. G., Hydrogenation of CO2 to Methanol: Importance of Metal-Oxide and Metal-Carbide Interfaces in the Activation of CO2. *Acs Catalysis* **2015**, *5*, 6696-6706.

(9) Rodriguez, J. A.; Grinter, D. C.; Liu, Z. Y.; Palomino, R. M.; Senanayake, S. D., Ceria-based model catalysts: fundamental studies on the importance of the metal-ceria interface in CO oxidation, the water-gas shift, CO2 hydrogenation, and methane and alcohol reforming. *Chemical Society Reviews* **2017**, *46*, 1824-1841.

(10) Del Vitto, A.; Sousa, C.; Illas, F.; Pacchioni, G., Optical properties of Cu nanoclusters supported on MgO(100). *J. Chem. Phys.* **2004**, *121*, 7457-7466.

(11) Di Valentin, C.; Giordano, L.; Pacchioni, G.; Rosch, N., Nucleation and growth of Ni clusters on regular sites and F centers on the MgO(001) surface. *Surf. Sci.* **2003**, *522*, 175-184.

(12) (a) Giordano, L.; Di Valentin, C.; Goniakowski, J.; Pacchioni, G., Nucleation of Pd dimers at defect sites of the MgO(100) surface. *Phys. Rev. Lett.* **2004**, *92*;   (b) Abbet, S.; Sanchez, A.; Heiz, U.; Schneider, W. D.; Ferrari, A. M.; Pacchioni, G.; Rosch, N., Acetylene cyclotrimerization on supported size-selected $Pd_n$ clusters ($1 <= n <= 30$): one atom is enough! *J. Am. Chem. Soc.* **2000**, *122*, 3453-3457;   (c) Giordano, L.; Goniakowski, J.; Pacchioni, G., Characteristics of Pd adsorption on the MgO(100) surface: Role of oxygen vacancies. *Phys. Rev. B* **2001**, *64*.

(13) Ferrari, A. M.; Xiao, C. Y.; Neyman, K. M.; Pacchioni, G.; Rosch, N., Pd and Ag dimers and tetramers adsorbed at the MgO(001) surface: a density functional study. *Phys. Chem. Chem. Phys.* **1999**, *1*, 4655-4661.

(14) Judai, K.; Abbet, S.; Worz, A. S.; Heiz, U.; Giordano, L.; Pacchiono, G., Interaction of Ag, Rh, and Pd atoms with MgO thin films studied by the CO probe molecule. *J. Phys. Chem. B* **2003**, *107*, 9377-9387.

(15) Giordano, L.; Pacchioni, G.; Ferrari, A. M.; Illas, F.; Rosch, N., Electronic structure and magnetic moments of $Co_4$ and $Ni_4$ clusters supported on the MgO(001) surface. *Surf. Sci.* **2001**, *473*, 213-226.

(16) Pacchioni, G.; Sicolo, S.; Di Valentin, C.; Chiesa, M.; Giamello, E., A route toward the generation of thermally stable Au cluster anions supported on the MgO surface. *Journal of the American Chemical Society* **2008**, *130*, 8690-8695.

(17) Lian, J. C.; Finazzi, E.; Di Valentin, C.; Risse, T.; Gao, H. J.; Pacchioni, G.; Freund, H. J., Li atoms deposited on single crystalline MgO(001) surface. A combined experimental and theoretical study. *Chemical Physics Letters* **2008**, *450*, 308-311.

(18) Chiesa, M.; Giamello, E.; Di Valentin, C.; Pacchioni, G.; Sojka, Z.; Van Doorslaer, S., Nature of the chemical bond between metal atoms and oxide surfaces: New evidences from spin density studies of K atoms on alkaline earth oxides. *Journal of the American Chemical Society* **2005**, *127*, 16935-16944.

(19) (a) Abbet, S.; Sanchez, A.; Heiz, U.; Schneider, W. D., Tuning the selectivity of acetylene polymerization atom by atom. *J. Catal.* **2001**, *198*, 122-127;   (b) Judai, K.; Worz, A. S.; Abbet, S.; Antonietti, J.-M.; Heiz, U.; Del Vitto, A.; Giordano, L.; Pacchioni, G., Acetylene trimerization on Ag, Pd and Rh atoms deposited on MgO thin films. *Physical Chemistry Chemical Physics* **2005**, *7*, 955-962.

(20) Abbet, S.; Riedo, E.; Brune, H.; Heiz, U.; Ferrari, A. M.; Giordano, L.; Pacchioni, G., Identification of defect sites on MgO(100) thin films by decoration with Pd atoms and studying CO adsorption properties. *Journal of the American Chemical Society* **2001**, *123*, 6172-6178.

(21) Simic-Milosevic, V., et al., Charge-induced formation of linear Au clusters on thin MgO films: Scanning tunneling microscopy and density-functional theory study. *Physical Review B* **2008**, *78*, 6.

(22) (a) Giordano, L.; Martinez, U.; Sicolo, S.; Pacchioni, G., Observable consequences of formation of Au anions from deposition of Au atoms on ultrathin oxide films. *J. Chem. Phys.* **2007**, *127*;   (b) Giordano, L.; Pacchioni, G., Charge transfers at metal/oxide interfaces: a DFT study of formation of K delta+ and Au delta- species on MgO/Ag(100) ultra-thin films from deposition of neutral atoms. *Phys. Chem. Chem. Phys.* **2006**, *8*, 3335-3341.

(23) Ricci, D.; Bongiorno, A.; Pacchioni, G.; Landman, U., Bonding trends and dimensionality crossover of gold nanoclusters on metal-supported MgO thin films. *Phys. Rev. Lett.* **2006**, *97*.

(24) Simic-Milosevic, V.; Heyde, M.; Nilius, N.; Konig, T.; Rust, H. P.; Sterrer, M.; Risse, T.; Freund, H. J.; Giordano, L.; Pacchioni, G., Au dimers on thin MgO(001) films: Flat and charged or upright and neutral? *Journal of the American Chemical Society* **2008**, *130*, 7814-+.

(25) Martinez, U.; Giordano, L.; Pacchioni, G., Tuning the work function of ultrathin oxide films on metals by adsorption of alkali atoms. *Journal of Chemical Physics* **2008**, *128*.

(26) Finazzi, E.; Di Valentin, C.; Pacchioni, G.; Chiesa, M.; Giamello, E.; Gao, H. J.; Lian, J. C.; Risse, T.; Freund, H. J., Properties of alkali metal atoms deposited on a MgO surface: A systematic experimental and theoretical study. *Chem.-Eur. J.* **2008**, *14*, 4404-4414.

(27) Smerieri, M.; Pal, J.; Savio, L.; Vattuone, L.; Ferrando, R.; Tosoni, S.; Giordano, L.; Pacchioni, G.; Rocca, M., Spontaneous Oxidation of Ni Nanoclusters on MgO Monolayers Induced by Segregation of Interfacial Oxygen. *J. Phys. Chem. Lett.* **2015**, *6*, 3104-3109.

(28) Tosoni, S.; Spinnato, D.; Pacchioni, G., DFT Study of $CO_2$ Activation on Doped and Ultrathin MgO Films. *J. Phys. Chem. C* **2015**, *119*, 27594-27602.

(29) (a) Chen, H. Y. T.; Giordano, L.; Pacchioni, G., From Heterolytic to Homolytic H-2 Dissociation on Nanostructured MgO(001) Films As a Function of the Metal Support. *J. Phys. Chem. C* **2013**, *117*, 10623-10629;   (b) Song, Z.; Xu, H., Unusual dissociative adsorption of H2 over stoichiometric MgO thin film supported on molybdenum. *Applied Surface Science* **2016**, *366*, 166-172;   (c) Song, Z.; Hu, H.; Xu, H.; Li, Y.; Cheng, P.; Zhao, B., Heterolytic dissociative adsorption state of dihydrogen favored by interfacial defects. *Applied Surface Science* **2018**, *433*, 862-868.

(30) (a) Di Valentin, C.; Del Vitto, A.; Pacchioni, G.; Abbet, S.; Worz, A. S.; Judai, K.; Heiz, U., Chemisorption and reactivity of methanol on MgO thin films. *J. Phys. Chem. B* **2002**, *106*, 11961-11969;   (b) Song, Z. J.; Xu, H., Splitting Methanol on Ultra-Thin MgO(100) Films Deposited on a Mo Substrate. *Phys. Chem. Chem. Phys.* **2017**, *19*, 7245-7251.

(31) Di Valentin, C.; Pacchioni, G.; Abbet, S.; Heiz, U., Conversion of NO to N2O on MgO thin films. *Journal of Physical Chemistry B* **2002**, *106*, 7666-7673.

(32) Di Valentin, C.; Pacchioni, G.; Chiesa, M.; Giamello, E.; Abbet, S.; Heiz, U., NO

monomers on MgO powders and thin films. *Journal of Physical Chemistry B* **2002**, *106*, 1637-1645.

(33) Song, Z.; Zhao, B.; Xu, H.; Cheng, P., Remarkably Strong Chemisorption of Nitric Oxide on Insulating Oxide Films Promoted by Hybrid Structure. *J. Phys. Chem. C* **2017**, *121*, 21482-21490.

(34) (a) Sambur, J. B.; Chen, T. Y.; Choudhary, E.; Chen, G. Q.; Nissen, E. J.; Thomas, E. M.; Zou, N. M.; Chen, P., Sub-particle reaction and photocurrent mapping to optimize catalyst-modified photoanodes. *Nature* **2016**, *530*, 77-80;   (b) Barth, C.; Reichling, M., Imaging the atomic arrangements on the high-temperature reconstructed alpha-Al2O3(0001) surface. *Nature* **2001**, *414*, 54-57;   (c) Zou, Z. G.; Ye, J. H.; Sayama, K.; Arakawa, H., Direct splitting of water under visible light irradiation with an oxide semiconductor photocatalyst. *Nature* **2001**, *414*, 625-627;   (d) Chueh, W. C.; Falter, C.; Abbott, M.; Scipio, D.; Furler, P.; Haile, S. M.; Steinfeld, A., High-Flux Solar-Driven Thermochemical Dissociation of CO2 and H2O Using Nonstoichiometric Ceria. *Science* **2010**, *330*, 1797-1801;   (e) Merte, L. R., et al., Water-Mediated Proton Hopping on an Iron Oxide Surface. *Science* **2012**, *336*, 889-893.

(35) Shin, H. J.; Jung, J.; Motobayashi, K.; Yanagisawa, S.; Morikawa, Y.; Kim, Y.; Kawai, M., State-selective dissociation of a single water molecule on an ultrathin MgO film. *Nature Materials* **2010**, *9*, 442-447.

(36) Jung, J.; Shin, H. J.; Kim, Y.; Kawai, M., Controlling water dissociation on an ultrathin MgO film by tuning film thickness. *Physical Review B* **2010**, *82*.

(37) Song, Z. J.; Fan, J.; Xu, H., Strain-induced water dissociation on supported ultrathin oxide films. *Sci. Rep.* **2016**, *6*.

(38) Song, Z. J.; Fan, J.; Shan, Y. Y.; Ng, A. M. C.; Xu, H., Generation of Highly Reactive Oxygen Species on Metal-Supported MgO(100) Thin Films. *Phys. Chem. Chem. Phys.* **2016**, *18*, 25373-25379.

(39) Altieri, S.; Contri, S. F.; Valeri, S., Hydrolysis at MgO(100)/Ag(100) oxide-metal interfaces studied by O 1s x-ray photoelectron and MgKL23L23 Auger spectroscopy. *Physical Review B* **2007**, *76*.

(40) (a) Jung, J.; Shin, H. J.; Kim, Y.; Kawai, M., Activation of Ultrathin Oxide Films for Chemical Reaction by Interface Defects. *Journal of the American Chemical Society* **2011**, *133*, 6142-6145;   (b) Jung, J.; Shin, H. J.; Kim, Y.; Kawai, M., Ligand Field Effect at Oxiide-Metal Interface on the Chemical Reactivity of Ultrathin Oxide Film Surface. *Journal of the American Chemical Society* **2012**, *134*, 10554-10561.

(41) Johnson, M. A.; Stefanovich, E. V.; Truong, T. N.; Gunster, J.; Goodman, D. W., Dissociation of water at the MgO(100)-water interface: Comparison of theory with experiment. *Journal of Physical Chemistry B* **1999**, *103*, 3391-3398.

(42) Gunster, J.; Liu, G.; Stultz, J.; Krischok, S.; Goodman, D. W., Water and methanol adsorption on MgO(100)/Mo(100) studied by electron spectroscopies and thermal programmed desorption. *Journal of Physical Chemistry B* **2000**, *104*, 5738-5743.

(43) Pacchioni, G.; Freund, H., Electron Transfer at Oxide Surfaces. The MgO Paradigm: from Defects to Ultrathin Films. *Chemical Reviews* **2013**, *113*, 4035-4072.

(44) Sterrer, M.; Risse, T.; Pozzoni, U. M.; Giordano, L.; Heyde, M.; Rust, H. P.; Pacchioni, G.; Freund, H. J., Control of the charge state of metal atoms on thin MgO films. *Physical Review Letters* **2007**, *98*.

(45) Perdew, J. P.; Burke, K.; Ernzerhof, M., Generalized Gradient Approximation Made Simple. *Phys. Rev. Lett.* **1996**, *77*, 3865-3868.

(46) (a) Grimme, S.; Antony, J.; Ehrlich, S.; Krieg, H., A consistent and accurate ab initio parametrization of density functional dispersion correction (DFT-D) for the 94 elements H-Pu. *J. Chem. Phys.* **2010**, *132*;   (b) Grimme, S.; Ehrlich, S.; Goerigk, L., Effect of the Damping Function in Dispersion Corrected Density Functional Theory. *J. Comput. Chem.* **2011**, *32*, 1456-1465.

(47) Blöchl, P. E., Projector Augmented-Wave Method. *Phys. Rev. B* **1994**, *50*, 17953-17979.

(48) (a) Stewart, G. R.; Giorgi, A. L., Specific heat of bcc $Mo_{1-x}Tc_x$. *Physical Review B* **1979**, *19*, 5704-5710;   (b) Feng, Z.; Babu, V. S.; Zhao, J.; Seehra, M. S., Effect of magnetic dilution on magnetic ordering in $Ni_pMg_{1-p}O$. *J. Appl. Phys.* **1991**, *70*, 6161-6163.

(49) Henkelman, G.; Uberuaga, B. P.; Jónsson, H., A climbing image nudged elastic band method for finding saddle points and minimum energy paths. *J. Chem. Phys.* **2000**, *113*, 9901-9904.

(50) Kresse, G.; Furthmüller, J., Efficiency of ab-initio total energy calculations for metals and semiconductors using a plane-wave basis set. *Comput. Mater. Sci.* **1996**, *6*, 15-50.

(51) Henkelman, G.; Arnaldsson, A.; Jonsson, H., A fast and robust algorithm for Bader decomposition of charge density. *Comput. Mater. Sci.* **2006**, *36*, 354-360.

(52) Humphrey, W.; Dalke, A.; Schulten, K., VMD: Visual molecular dynamics. *J. Mol. Graph.* **1996**, *14*, 33-38.

(53) Wang, Y., *VASPMO, version 0.31*, https://sourceforge.net/projects/vaspmo/ (accessed Oct 24, 2016).

(54) Momma, K.; Izumi, F., VESTA: a three-dimensional visualization system for electronic and structural analysis. *J. Appl. Crystallogr.* **2008**, *41*, 653-658.